\documentclass[12pt,a4paper]{article}

\usepackage{amsmath}

\usepackage{amsfonts}
\usepackage{amssymb}
\usepackage{color}

\begin{document}

\title{Testing boundaries of applicability of quantum probabilistic formalism to modeling of cognition}

\author{ Irina Basieva, Andrei Khrennikov\\
 International Center for Mathematical Modeling \\in Physics and Cognitive Science \\
Linnaeus University,  V\"axj\"o, Sweden }

\maketitle

\begin{abstract}
Recently the mathematical formalism of quantum mechanics, especially methods of quantum probability theory, started to be widely used in a variety of applications 
outside of physics, e.g., cognition and psychology as well as economy and finances.  To distinguish such models from genuine quantum physical models, they often called quantum-like
(although often people simply speak about, e.g., ``quantum cognition''). These novel applications generate a number of foundational questions. Nowadays 
we can speak about a new  science - foundations of quantum-like modeling. At the first stage this science was mainly about comparison of classical and quantum models, mainly 
in the probabilistic setting. It was found that statistical data from cognitive psychology violate some basic constraints posed on data by classical probability theory (Kolmogorov, 1933); 
in particular, the constraints given by the formula of total probability and Bell's type inequalities. Recently another question attracted some attention. In spite of real success in applications,
there are no reason to believe that the quantum probability would cover completely all problems of, e.g., cognition. May be more general probability models have to be explored. 
A similar problem attracted a lot of attention in foundations of quantum  physics culminating in a series of experiments to check Sorkin's equality for the triple-slit experiment by 
Weihs' group. In this note we present a similar test in the cognitive experimental setting. Performance of this test would either give further confirmation of the adequacy    of 
the quantum probability model to cognitive applications or rejection of the conventional quantum model. Thus this note opens the door for a series of exciting experimental tests
for the quantum-like model of cognition. 
\end{abstract}

\section{Introduction}

Recently nonclassical (``non-Kolmogorovean'') probabilistic structure of experimental statistical data collected in cognitive science, psychology, decision making was successfully modeled with the aid 
of quantum probability theory, e.g., open access  review  \cite{KHR_FRONT}  and monograph \cite{KHR_UB}
and the extended lists of references in them.\footnote{The idea that quantum and cognitive have something in common is not new. However, before 
it was presented either in philosophic discussions, e.g., between Jung and Pauli (also Whitehead) or in abstract mathematical models, e.g., 
\cite{KHR0}. Nowadays numerous research groups work with real experimental data. We also point from very beginning that 
this activity is devoted to so to say quantum-like modeling cognition, i.e., it has to be sharply distinguished from attempts to model cognition on the basis 
of  genuine quantum physical processes in the brain - in the spirit of, e.g., Penrose. } The latter is based on representation of probabilities by squared 
complex amplitudes (Born's rule of quantum mechanics). Quantum calculus relaxes a few important classical probabilistic  constraints. For example, one of the basic
laws of classical probability theory, the formula of total probability (FTP) is violated. In physical terms violation of FTP for quantum observables 
is represented as interference of probabilities.  As was found found by one of the authors (AKH), see \cite{KHR_FRONT}, FTP is violated as well for statistical data collected in cognitive psychology. In this way 
one can model the  disjunction effect. Another important cognitive effect, the order effect, can be modeled by exploring noncommutativity of observables, see
also \cite{KHR_FRONT} for review on modeling of variety of probability fallacies and paradoxes, e.g., Elsberg paradox. 

In principle there are no reasons to expect that quantum probability matches perfectly  cognition. Yes, we are sure that cognition cannot be represented with the aid 
of classical probability (Kolmogorov's set-theoretic axiomatics, 1933) , e.g., because FTP is violated, see also \cite{KHR_FRONT} for violation of Bell inequality (another important 
classical probabilistic constraint). However, it may happen that quantum probability, although so successful for modeling of some features of cognition, cannot cover 
completely cognitive phenomena. There might be some effects which (probabilistically) are neither classical nor quantum. Recently this problem attracted some attention \cite{KHR_UB},
\cite{PLOS}-\cite{eric2}. 

We remark that there are neither reasons to expect that quantum probability covers all statistical experiments in quantum physics. The mathematical formalism 
of quantum mechanics was not derived from some physically natural principles as, e.g., Einstein's principle of relativity.\footnote{Nowadays tremendous efforts are 
put to find such principles and derive the quantum formalism from them, see, e.g., \cite{Z0}-\cite{BR4}. However, it is too early to consider this important project as completed.} 
It was recognized as a fruitful theory, because it was successfully applied to a number of fundamental problems of physics of micro-world. 
However, we cannot exclude that one day physicists would find that 
statistical data from some experiment cannot be described by quantum probability theory. If it were happened, a new calculus of probabilities should be designed.

Recently one of world's best experimental groups working in quantum foundations, the group of prof. G. Weihs from Innsbruck, put tremendous  efforts to find 
violations of the basic rule of quantum probability theory, the Born rule, by testing Sorkin's equality   \cite{Sorkin0}, see \cite{Sorkin1}, \cite{Sorkin2}.\footnote{Another model leading to violation of this rule  was proposed  in 
the framework of so-called {\it prequantum classical statistical field theory} (PCSFT) \cite{Beyond}. In PCSFT,  Born' rule is perturbed, 
quantum probability is defined as $p(x)= \vert \psi(x) \vert^2 + \alpha \vert \psi(x) \vert^4, \alpha <<1.$ Some experimental design to check  this modification of Born's rule 
was proposed  \cite{Beyond} (but it has not yet been performed). We remark that, in contrast to Sorkin's formal probabilistic reasoning, the PCSFT-modification of Born's is based
on physical (classical random field) argument. Thus, in some sense, this model is more interesting from the physical viewpoint.  On the other hand, the abstract probabilistic realization 
serves perfectly to our present aim - applications outside of physics.} One can design a simple test of its validity (based on the work of R. Sorkin \cite{Sorkin0}) 
in the form of the triple-slit experiment.
This is really surprising, because the two-slit experiment is one of the most fundamental experiments confirming quantum probability calculus, see Feynman et al. 
 \cite{Feynman}. It was done by numerous
research groups for a variety of quantum systems. And by adding just one slit it is possible to check the validity of quantum theory. Unfortunately, simplicity on 
the theoretical level confronts extreme complexity on the experimental level. Preliminary it was announced that in the triple-slit experiment 
violation of Born's rule was not found.\footnote{However, it is not easy to estimate correctly the impact of such technicalities as nonlinearity of photo-detectors and 
experiments (struggling with nonlinearity of detectors) are continued.} 

Now a few words about the test. Sorkin found \cite{Sorkin0} that, although quantum probabilities violate FTP (or in other terms - additivity of probability), they satisfy some 
equality for disjunction of three  ``quantum events''.  In the experiment with slits, the terms of this equality correspond to the following experimental contexts: 
$C_{123}$ - all three slits are open, $C_{ij}, i \not=j,$ - only  slits $(i,j)$ are open, and $C_{i}$  - only  $i$th slit is open. Quantum theory predicts that statistical 
data collected for these (incompatible) experimental contexts has to satisfy Sorkin's equality. If its violation  were found, it would mean that from the probabilistic viewpoint
quantum theory is even more exotic. There also exist Sorkin's equalities of higher orders. And Weihs' group also works to check the fifth order interference (the talk of Gregor Weihs 
at V\"axj\"o-14, 15 conferences).   
Sorkin's equality corresponding to the $n$-slit experiment will be called Sorkin's $n$-equality.

This triple-slit test on the probabilistic structure of data can be designed for cognitive experiments. It seems (at least before starting experimenting) that it is easier 
to perform than the corresponding test with quantum systems - the process of detection is straightforward. And here we have higher expectation of violation 
of Born's rule than in physics, cf. with the experiment discussed in \cite{PLOS}, see also \cite{eric1}, \cite{eric2} for discussions.
In this note we restrict our consideration to Sorkin's $3$-equality, but 
it is easy to generalize our scheme of testing of the probability structure of cognition to the case of arbitrary $n.$

Both Sorkin \cite{Sorkin0} and Weihs et al. \cite{Sorkin1}, \cite{Sorkin2} worked with wave functions and, although the output of their calculations is correct, some additional efforts have to be put in 
the rigorous mathematical presentation. The latter is based on theory of quantum conditional probabilities and it will be done in this paper, see appendix.

\section {Sorkin's inequality}
\label{S}

In classical measure-theoretic framework, for two disjoint events $A_1$ and $A_2$ and an event $B$ we have
\begin{equation}
\label{CL1}  
p(B\wedge (A_1 \vee A_2)= p(B\wedge A_1) + p(B\wedge  A_2).
\end{equation}
This is the basic feature of classical probability, its additivity.
As was pointed out by Feynman, quantum probability is not additive \cite{Feynman}.  Formally, this is correct. However, one has to be careful 
in defining  the probability of conjunction of ``quantum events'', because they can be incompatible.  It seems that the only way to define rigorously 
such probability is to use quantum conditional probability which is well defined. Thus we want to explore a quantum analog of the classical equality:
\begin{equation}
\label{CL2}  
p(B\wedge A)= p(A) p(B\vert  A).
\end{equation}
This equality is a consequence of Bayes' formula
\begin{equation}
\label{CL3}  
 p(B\vert  A)= p(B \wedge  A)/ p(A).
\end{equation}
The latter is the definition of conditional probability in the Kolmogorov model. Thus in classical probability we start with well defined probability of conjunction of events, 
the joint probability, and then define conditional probability. In quantum probability we proceed another way around. We shall start with conditional probability and the define
joint probability. Taking into account the fundamental role which is played by conditional probability in our further considerations, it is useful to rewrite equality (\ref{CL1}), additivity law,
in terms of conditional probabilities:   
\begin{equation}
\label{CL1c}  
p(B \vert A_1 \vee A_2)= \frac{1}{p(A_1 \vee A_2)}[p(A_1) p(B\vert A_1) +  p(A_2) p(B\vert  A_2)].
\end{equation}
This is one of the basic elements of classical statistical inference, {\it the formula of total probability.} In particular, if $p(A_1 \vee A_2)=1,$ we get the formula of total probability in its 
the most  commonly used form: 
\begin{equation}
\label{CL2c}  
p(B)= p(A_1) p(B\vert A_1) +  p(A_2) p(B\vert  A_2).
\end{equation}
Preparing to quantum considerations, let us introduce the {\it ``interference term''}:
\begin{equation}
\label{CL2ca}  
I_{12}= p(A_1 \vee A_2) p(B \vert A_1 \vee A_2)  - p(A_1) p(B\vert A_1) -  p(A_2) p(B\vert  A_2)
\end{equation}
\[
 = p(B\wedge (A_1 \vee A_2)) -  p(B\wedge A_1) - p(B\wedge  A_2).
\]
In classical probability theory $I_{12}=0$ (but in quantum theory $I_{12}\not=0,$ see appendix, equation (\ref{3LM})). 

Now consider three events $A_i, i=1,2,3.$ Let  $p(A_1 \vee A_2 \vee A_3)=1.$ Here the additivity law  gives us  
\begin{equation}
\label{CL3c}  
p(B)= p(B\wedge (A_1 \vee A_2 \vee A_2)= p(B\wedge A_1) + p(B\wedge  A_2) + p(B\wedge A_3).
\end{equation}
And the formula of total probability has the form: 
\begin{equation}
\label{CL4c}  
p(B)=  p(A_1) p(B\vert A_1) + p(A_2) p(B\vert  A_2) + p(A_3) p(B\vert A_3).
\end{equation}
We introduce the corresponding ``interference coefficient'': 
\begin{equation}
\label{CL5c}  
I_{123}=  p(B) -  p(A_1) p(B\vert A_1) + p(A_2) p(B\vert  A_2) + p(A_3) p(B\vert A_3) 
\end{equation}
\[
= p(B) - p(B\wedge A_1) - p(B\wedge  A_2) + p(B\wedge A_3).
\]
In classical probability theory $I_{123}=0.$ 

Since $I_{12}=I_{13}= I_{23}=0,$ we can write this term as
\begin{equation}
\label{CL5c}  
I_{123}=  p(B) -  p(A_1) p(B\vert A_1) -  p(A_2) p(B\vert  A_2)  -  p(A_3) p(B\vert A_3) 
\end{equation}
\[
- I_{12} - I_{13} - I_{23}.
\]
Thus 
\begin{equation}
\label{CL6c}  
I_{123}=  p(B) -  p(A_1) p(B\vert A_1) - p(A_2) p(B\vert  A_2) - p(A_3) p(B\vert A_3) 
\end{equation}
\[
- p(A_1 \vee A_2) p(B \vert A_1 \vee A_2)  + p(A_1) p(B\vert A_1)  +  p(A_2) p(B\vert  A_2) +....
\]
Finally, we obtain the  triple-interference coefficient in the following form: 
  \begin{equation}
  \label{CL7c}  
 I_{123}=  p(B) -  p(A_1 \vee A_2) p(B \vert A_1 \vee A_2) - p(A_1 \vee A_3) p(B \vert A_1 \vee A_3) - p(A_2 \vee A_3) p(B \vert A_2 \vee A_3)
  \end{equation}
\[
+ p(A_1) p(B\vert A_1)  +  p(A_2) p(B\vert  A_2)  + p(A_3) p(B\vert A_3) .
\]
By using the joint probability distribution and by shortening notation, $p_{ij}= p(B\wedge (A_i \vee A_j),  p_{i}= p(B\wedge  A_i ),,$ we write this coefficient as
\begin{equation}
\label{CL8c}  
I_{123} =  p_{123} - p_{12}- p_{13} - p_{23} + p_1 + p_2 + p_3.
\end{equation}
Of course, in classical probability the expression in the left-hand side also equals to zero. 
Surprisingly the coefficient defined by the left-hand side of  (\ref{CL8c}) also equals to zero, 
in spite non-vanishing (in general) of $I_{ij}.$ And this was an interesting discovery of R. Sorkin \cite{Sorkin0}.
We shall prove this in appendix by proceeding in the rigorous framework of quantum conditional probabilities, see equality (\ref{3L}).
(In principle, we can proceed with only conditional probabilities, i.e., without joint probabilities at all. However, Sorkin formulated 
his equalities in terms of joint probabilities and we wanted two have similar expressions.)

Now we can forget about quantum probabilities and just to check whether some statistical data satisfies Sorkin's equality 
\begin{equation}
\label{CL11c}  
I_{123} =  p_{123} - p_{12}- p_{13} - p_{23} + p_1 + p_2 + p_3= 0
\end{equation}
 or not.
If Sorkin's equality were violated, both classical and quantum models should be rejected; in the opposite case, we would get another (nontrivial)
confirmation of validity of the quantum model. 

In coming experiments, instead of one event $B,$ we shall consider a few disjoint events $B_j.$ We shall use this $j$ as the upper index for probabilities, e.g., 
$p_i^{(j)}.$

\section{Experiment}

We present a toy model of  the cognitive analog of the triple-slit experiment. 
There is a homogeneous group of people recruited for the experiment.\footnote{Its homogeneity  is important, because 
it will be divide into a few subgroups which will be used to collect different blocks of statistical data. And it is 
important that we  can assume that the members of all subgroups have ``the same mental state''.}
They are informed that during the experiment they will answer to a few questions related to their possible emigration to other 
countries; for this experiment, three fixed countries; for example, we can select Brazil, Canada, Australia, $a=1,2,3.$

We tell them the story: ``Suppose you would like to emigrate to one of these countries.'' 
Then this group if divided into three subgroups $G, G^\prime$ and $G^{\prime \prime}.$ People from the first two groups will 
participate in experiments with conditional questioning and in the last group in unconditional experiment.  

Those from $G$ first are asked the $a$-question: {\it To which of these three countries would you  like to emigrate?}
Those with the answers $a=i$ form the new groups $G_i.$ We find the probabilities $p(a=i) \approx \frac{n_{G_i}}{n_{G}},$
where $n_Q$ denotes the number of elements in the set $Q.$ 

Then we have to ask those in groups $G_i$ another question, say $b,$ which has to 
 be ``complementary'' to the $a$-question. Selection of $b$ is the delicate issue.  
 For example, let us proceed with the question $b:$ {\it Are you ready to change your profession?} (in the case of emigration to this country)\footnote{Another proposal: {\it Do you think that 
your application for emigration (to this  country)  will be successful?}}
 This is the dichotomous observable $b=0, 1$ corresponding to the answers   `no', `yes'. (In principle, we can consider $b$ having any finite number 
of values, but it is enough to find violation of Sorkin's equality for a dichotomous $b$-observable.)
Those in $G_i$ who answered $b=j$ form the group denoted by $G_{j\vert i}.$ Now we can find conditional probabilities,
$p(b=j\vert a=i)\approx \frac{n_{G_{j\vert i}}}{n_{G_{i}}}$ and the ``ordered joint probabilities''
$p_{i}^{(j)} = p(a=i) p(b=j\vert a=i) \approx \frac{n_{G_{j\vert i}}}{n_{G}}.$

People from the group $G^\prime$ are asked about pairs of countries:  {\it Which pair of these three countries would you select to emigrate?}
The answers are pairs $(k, m).$ The groups  corresponding to concrete pairs are denoted as $G^\prime_{km}.$ And we can find the probabilities
$p(a=k \vee a= m) \approx \frac{n_{G^\prime_{km}}}{n_{G^\prime}}.$ Then we ask those in each group the $b$-question; depending on the answer 
$b=j,$ we form the groups $G^\prime_{j\vert km}.$ and find the conditional 
probabilities $p(b=j \vert   a=k \vee a= m) \approx \frac{n_{G^\prime_{j \vert km}}}{n_{G^\prime_{km}}}.$ They determine the joint probabilities
$p^{(j)}_{km}= p(a=k \vee a= m)  p(b=j \vert   a=k \vee a= m) \approx \frac{n_{G^\prime_{j \vert km}}}{n_{G^\prime}}.$  

Now those  in $G^{\prime \prime}$ are asked just the $b$-question; depending to the answers we find 
the probabilities $p(b=j)= p^{(j)}_{123}.$ 

Finally, we put collected probabilities into Sorkin's equality, to check whether the interference term $I_{123}^{(j)}=0.$ 

Of course,  it is useful before to start this ``triple-slit experiment'', to check whether the questions are really complementary, i.e.,
one has to start  with the corresponding two slit versions of this experiment to see whether $I_{km}^{(j)}\not=0.$ (But for this experiment 
a new group of people has to be used.) 

\section*{Appendix: Derivation of Sorkin's equality of the third order}

We restrict our consideration by standard quantum observables given by Hermitian operators $a$ and $b.$ We consider the finite-dimensional case.
Unfortunately, we have to assume that their eigenvalues can be degenerate, because as was found in \cite{KHR_UB} (see also \cite{eric1}, \cite{eric2} for detailed analysis),  
it is impossible to represent some cognitive entities by operators with nondegenerate spectra.\footnote{Such operators have to 
generate double-stochastic matrices of transition probabilities. However, the real data from cognitive psychology do not satisfy to this constraint
(not clear why...). }     

Let  $a_i, i=1,2,3,$ and $b_j, j=1,2,...,m,$ be the eigenvalues of $a$ and $b.$ Denote the corresponding projectors by $P^a_i$ and $P^b_j$ respectively.
We shall also introduce projector $P^a_{ik}, i\not= k,$ on subspaces consisting of eigenvectors of $a$ corresponding to eigenvalues $\alpha_i$ and $\alpha_k.$ 
Thus $P^a_{ik} = P^a_{i} + P^a_{k}.$

Let $\rho$ be a  quantum state. Then we have:
$
p(a=a_i) = {\rm Tr}  P^a_i  \rho P^a_i,
p(a=a_k \vee a=a_m)= {\rm Tr}  P^a_{km}   \rho P^a_{km}.
$
By definition of conditional probability in quantum probability theory 
$$   
p(b= b_j\vert a=a_i)=  \frac{{\rm Tr} P^b_j P^a_i  \rho P^a_i}{{\rm Tr}  P^a_i  \rho P^a_i}.  
$$
We consider also ``ordered joint probability distribution" 
$
p_i \equiv  p(a=a_i) p(b= \beta_j\vert a=a_i),
$
the index $j$ is fixed and we omit it.  For $k\not=m,$ we consider the two-slit interference: 
$$
p(b= b_j\vert a=a_k \vee a=a_m)= \frac{{\rm Tr} P^b_j P^a_{km}  \rho P^a_{km} }{{\rm Tr}  P^a_{km}   \rho P^a_{km}}=
$$
$$
=\frac{1}{{\rm Tr}  P^a_{km}   \rho P^a_{km}} \Big({\rm Tr}  P^a_k  \rho P^a_k \frac{{\rm Tr} P^b_j P^a_k  \rho P^a_k}{{\rm Tr}  P^a_k  \rho P^a_k} + 
{\rm Tr}  P^a_m  \rho P^a_m \frac{{\rm Tr} P^b_j P^a_m  \rho P^a_m}{{\rm Tr}  P^a_m  \rho P^a_m} +  
+ I_{km} \Big).
$$
where $I_{km} = {\rm Tr} P^b_j P^a_k  \rho P^a_m +  {\rm Tr} P^b_j P^a_m  \rho P^a_k.$
We consider also ``ordered joint probability distribution"
$
p_{km} \equiv p(a=a_k \vee a=a_m) p(b= b_j\vert a=a_k \vee a=a_m).
$ 
Thus 
\begin{equation}
\label{3L}
p_{km}  =  p_{k} +  p_{m} + I_{km}.   
\end{equation}
This is the quantum modification of the additivity law; in fact, this is the quantum  analog of FTP, since joint probabilities are defined via conditional probabilities.
R. Feynman emphasized \cite{Feynman} non-additivity of quantum probability; by using the language of conditional quantum probabilities one of the authors of this paper 
reformulated this violation as disturbance of FTP \cite{KHR_FRONT}, \cite{KHR_UB}. Thus in quantum theory we have that in general
\begin{equation}
\label{3LM}
 I_{km}= p_{km}  -  p_{k} -  p_{m} \not=0. 
\end{equation}

Now we consider the triple-slit interference. We shall use the equality $I= \sum_i P^a_i.$  We have:  
$$
p_{123}\equiv p(b= \beta_j)= p(b= \beta_j \vert a=a_1 \vee a=a_2 \vee a= a_3)  ={\rm Tr} P^b_j  \rho 
$$
$$
= \sum_i {\rm Tr}  P^a_i  \rho P^a_i \frac{{\rm Tr} P^b_j P^a_i  \rho P^a_i}{{\rm Tr}  P^a_i  \rho P^a_i}  
$$
$$
({\rm Tr} P^b_j P^a_1  \rho P^a_2 +  {\rm Tr} P^b_j P^a_2  \rho P^a_1) + ({\rm Tr} P^b_j P^a_1  \rho P^a_3 + {\rm Tr} P^b_j P^a_3  \rho P^a_1) +
({\rm Tr} P^b_j P^a_2  \rho P^a_3+ {\rm Tr} P^b_j P^a_3  \rho P^a_2).
$$
 $$
 = p_1 + p_2 + p_3  + I_{12} + I_{13} + I_{23}=
 $$
 $$
 = p_1 + p_2 + p_3  + 
(p_{12}  - p_{1} - p_{2}) + (p_{13}  - p_{1} - p_{3}) + (p_{23}  - p_{2} - p_{3})   
$$
Hence, as well as in classical probability theory,  see section \ref{S}, 
\begin{equation}
\label{3L}
I_{123} =  p_{123} - p_{12}- p_{13} - p_{23} + p_1 + p_2 + p_3 =0.
\end{equation}

\section*{Acknowledgments} 

One of the authors (AKH) would like to thank G. Weihs for numerous discussions on the possibility to violate Born's rule, in particular on the triple-slit experiment, and the possibility 
to see the lab and performance of this test during the visit to Innsbruck in May 2013 and hospitality during this visit.

\end{document}